\documentclass[epj]{svjour} %{svjour2}

\usepackage{color}
\usepackage{graphics}
\usepackage{amsmath}
\usepackage{amssymb}
\begin{document}
\title{Ion-impact-induced multifragmentation of liquid droplets}
\author{Eugene Surdutovich\inst{1,3} \and Alexey Verkhovtsev\inst{2,3}
    \and Andrey V. Solov'yov\inst{3,4}
}                     % Do not remove
\offprints{Eugene Surdutovich}  
\institute{Department of Physics, Oakland University, Rochester, Michigan 48309, USA
\and Instituto de F\'{\i}sica Fundamental, CSIC, Serrano 113-bis, 28006 Madrid, Spain
\and MBN Research Center,
% Altenhoferallee 3
60438 Frankfurt am Main, Germany
\and On leave from A.F. Ioffe Physical Technical Institute, % Politechnicheskaya 26,
194021 St. Petersburg, Russian Federation}
\date{Received: date / Revised version: date}
\abstract{An instability of a liquid droplet traversed by an energetic ion is explored. This instability is brought about by the predicted shock wave induced by the ion. An observation of multifragmentation of small droplets traversed by ions with high linear energy transfer is suggested to demonstrate the existence of shock waves. A number of effects are analysed in effort to find the conditions for such an experiment to be signifying. The presence of shock waves crucially affects the scenario of radiation damage with ions since the shock waves significantly contribute to the thermomechanical damage of biomolecules as well as the transport of reactive species. {While the scenario has been upheld by analyses of biological experiments, the shock waves have not yet been observed directly, regardless of a number of ideas of experiments to detect them were exchanged at conferences.}
} %end of abstract
\maketitle

\section{Introduction}

The Rayleigh instability at the liquid surface has been studied since 1880s for a number of different applications~\cite{Rayleigh,SirTaylor,BetheWeiz,Jellium,Lyalin}. The main aspect in all of these applications is that the integrity of a liquid object or an object that features certain liquid properties is related to the stability of its interface. If this stability is disturbed either by Coulomb repulsion or other forces, the object may disintegrate to form smaller more stable objects. This is equally applied to charged liquid droplets, unstable nebulae structure, heavy nuclei experiencing fission, and unstable clusters. In this work, we investigate a new type of instability of liquid droplets induced by energetic ion's passage. The mechanism that triggers such an instability is ascribed to the shock waves induced by ions propagating in a liquid medium~\cite{prehydro}. Moreover, the experimental observation of a multifragmentation of liquid droplets traversed by ions will be the most direct evidence of existence of shock waves. However, let us proceed in order.

{
Ion beams have been clinically used for radiotherapy since 1990s, with protons and carbon ions being the most used projectiles~\cite{FokasKraft09,SchardtRMP10}. The attractiveness of ions as projectiles compared to commonly used x-rays is in the existence of the Bragg peak at the end of the depth-dose curve. This peak occurs due to the increase of ionization cross section as the velocity of ions decreases. Its position for a given medium and projectile solely depends on the initial energy of ions and, therefore, can be focused to the tumour, while the surrounding regions acquire much smaller doses and less prone to radiation damage.

The optimisation of ion-beam therapy requires a thorough understanding of the relation between the physical properties of radiation with biological damage, which is quantified as a percentage of inactivated cells in the irradiated region. In a conventional x-ray radiotherapy the relation between the deposited dose and biodamage is understood on a semi-empirical level for many different cells and conditions~\cite{Hall,Alpen}. A staggering biological diversity that includes various cell repair mechanisms makes it difficult to construct a reliable predictive method of assessment of radiation damage. The situation with ion beams is different despite the lack of complete understanding of effects following the ion's traverse through tissue. Stronger physical and chemical effects in this case make the repair and other biological factors less important and thus increase the hope to achieve a quantitative assessment of radiation damage. Nevertheless, empirical methods of relating the dose deposited by ions to the biodamage appeared first~\cite{SchardtRMP10}.
%For example, it is understood that ions lose energy as they ionise or excite the molecules of the medium, and the ejected electrons and other formed reacting species may interact with biomolecules and thus damage cells.
The track-structure community is engaged in modeling of the transport and interactions of reactive species in ion's tracks leading to DNA damage by means of Monte Carlo simulations~\cite{Nikjoo06,StewartMCDS,FriedlandSR2017}. However, these methods do not account for all effects that involved in the scenario of radiation damage and, therefore, important for treatment planning. One of these effects is the formation of shock waves following the energy relaxation in the vicinity of the ions' paths. The onset of shock waves is predicted at the time just following the end of formation of the track structure and the latter can perhaps provide the most detailed initial conditions for this process.
}

The existence of such waves has been predicted in the process of investigation of physical effects relevant to the radiation damage with ions, specifically in relation to ion-beam therapy, summarised as a multiscale approach to the physics of radiation damage with ions (MSA)~\cite{MSAColl,pre,IBCTbook}. {A number of experimental observations, such as the detection of acoustic waves (which are likely to be the remnants of shock waves) initiated by ions' passage through a medium~\cite{YYSun,Baily} or recent exploration of cavitation in liquid water caused by x-ray pulses~\cite{WaterCavit}, makes the shock waves initiated by ions plausible.}  Presence of shock waves significantly affects the transport of reactive species (free radicals, peroxide, etc.), may directly rupture a biomole\-cule~\cite{natnuke}, and, hence, is important for the assessment of radiation biodamage, and, eventually, therapy planning. {A recent analysis of cell survival probabilities measured in different conditions~\cite{CellSurSR16} strongly supports the shock wave scenario.}
Still, even though a number of ideas of experiments to detect shock waves were exchanged at conferences, they have not been realised, and thus far, this effect has not been observed. In this paper, we are drawing attention of a broad collision-physics community to this important problem and suggest an idea of an experiment for observation of a shock wave induced by an ion propagating in the medium.

\section{The effect of ion-induced shock waves on the medium}

In Ref.~\cite{prehydro}, it has been predicted that a traverse of liquid water or similar medium by an ion with energy of 0.1-0.5~MeV/u, which guarantees a large linear energy transfer (LET), brings about a shock wave that propagates radially away from the ion's path. This shock wave forms because the ion deposits energy (mainly by ionizing the medium) within a thin cylinder adjacent to its path. This energy is relaxed inside this cylinder, referred to as a hot cylinder, since the energy transfer outside of this cylinder is too slow.  Indeed, in Ref.~\cite{preheat}, it has been estimated that the heat conductivity mechanisms are capable of transferring the energy out of hot cylinder by about 100~ps after ion's traverse. The heat transfer by diffusing molecules also happens on the 100-ps scale because of the relatively (to electrons) small diffusion coefficients. However, according to Ref.~\cite{Radicals} by as soon as 50~fs, the electron contribution to heat transfer terminates since by then they lose most of their energy, and a high pressure (up to 100~GPa) is developed within the hot cylinder. Such a pressure build-up constitutes a formation of the wave front and the onset of the cylindrically symmetric shock wave that would propagate radially until it weakens~\cite{vilnius,natnuke} and becomes acoustic.

The energy deposited inside the hot cylinder is described by the LET, which for ions in the energy range of 0.1-0.5~MeV/u (relevant to this work)
%vicinity of the Bragg peak (area close to the maximum of LET)
is similar to the stopping power, $S=-d{\cal E}/dx$, where ${\cal E}$ is the energy of the ion and $x$ is the longitudinal coordinate.  According to Refs.~\cite{prehydro,MSAColl}, the ``strength'' of the shock wave is defined by the part of the stopping power that is related to ionization and excitation processes, $S_e$.
%In the beginning, a strong explosion model consistently describes the wave; then it gradually diffuses~\cite{vilnius}.
The pressure on the front of the shock wave is expressed via the velocity of the wave front $u$ as $P=1/(\gamma+1)\varrho u^2$~\cite{prehydro,LL6}, which can be written in terms of $S_e$ as
\begin{eqnarray}
P(r)=\frac{1}{\gamma+1} \frac{\beta^4}{2}\frac{S_e}{r^2},
\label{perrsureSW}
\end{eqnarray}
where $\gamma=C_P/C_V\approx1.2$, $\beta=0.86$ is a dimensionless constant, $\varrho=1$~g~cm$^{-3}$ is the density of medium (liquid water), and $r$ is the radius of the wave front.

The stopping power for a non-relativistic ion can be estimated by the Bethe-Bloch formula~\cite{Bethe,Bloch1},
\begin{eqnarray}
S_e=\frac{4\pi n_e \alpha^2 (\hbar c)^2}{m_e}\frac{z^2}{v^2}\ln{\frac{2 m_e v^2}{\bar I}},\label{bethe}
\end{eqnarray}
where $ze$ and $v$ are the charge and velocity of the ion, $e$, $m_e$, and $n_e$ are the electron charge, mass and number density, $\alpha$ is the fine structure constant, and ${\bar I}$ is the average excitation/ionization energy of molecules of the medium. When energetic ions enter a tissue-like medium (as is in the case of ion therapy) the value of $S_e$ is typically less than 10~eV/nm (for 400-MeV/u carbon ions used in therapy); then $S_e$ increases by the factor of almost 90 in its maximum called the Bragg peak~\cite{epjd}. This maximum is achieved when ion's energy is below 0.5~MeV/u. However, if we are interested in irradiating droplets, the Bragg peak does not occur, since an ion propagates in vacuum before hitting a droplet and we, at least for now, do not consider multiple droplets hit by the same ion. Therefore, $S_e$ is just estimated by Eq.~(\ref{bethe}), and is (for a given droplet) a function of ion's charge and velocity.

After the high-LET ion's traverse, it is predicted that such a shock wave is formed when the initial pressure profile is developed according with the radial dose distribution~\cite{Radicals}. As the shock wave propagates in the radial direction away from the ion's path, it causes cavitation in its wake. Later on the pressure on the wave front weakens and the cylindrical cavity formed near the axis of the hot cylinder fills in. If an ion traverses a bulk medium (such as liquid water), it is expected that by the time of about 1~ns, only acoustic waves initiated by the above shock-rarefaction wave dynamics reveal the ion's passage and its vigorous interaction with the medium.

\section{Shock wave interaction with liquid droplets}

\subsection{Multifragmentation of a droplet due to a shock wave}

Now imagine a high-LET ion traversing a liquid spherical droplet of radius $R$. According to the predictions of Ref.~\cite{prehydro}, a shock wave propagates away from the ion's path.  When the shock wave front reaches the surface of the droplet, the pressure at the edge is given by~(\ref{perrsureSW}). As is well known, a spherical droplet is held up in its shape by the surface tension pressure equal to $2\sigma/R$, where $\sigma$ is the coefficient of surface tension. Therefore, if the pressure on the wave front is higher than that of surface tension, the droplet should explode. In order to make the most conservative estimate, let us consider the ion whose trajectory passes through the center of the droplet. Then, the condition for a droplet's explosion can be written as an inequality,
\begin{eqnarray}
\frac{1}{\gamma+1} \frac{\beta^4}{2}\frac{S_e}{R^2}>\frac{2\sigma}{R}. \label{condition}
\end{eqnarray}

\begin{figure}
\begin{centering}
\resizebox{0.99\columnwidth}{!}
{\includegraphics{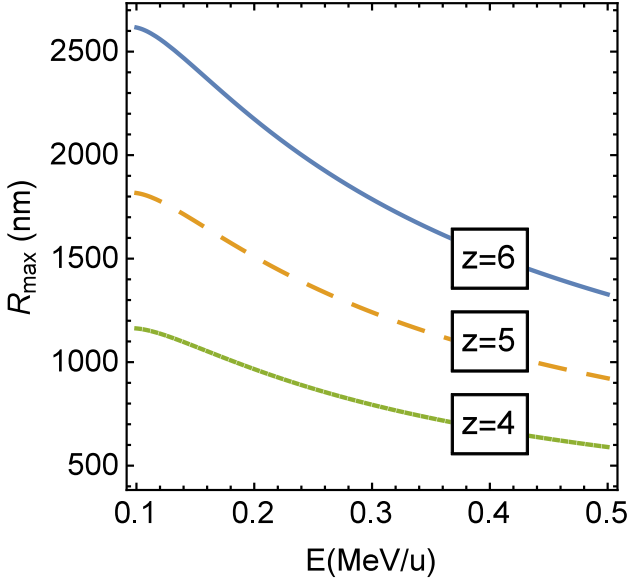}}
\caption{\label{fig:RE} The dependence of the maximum radius of the liquid water droplet (at room temperature), at which its explosion due to the effect of the ion-induced shock wave is predicted, on the energy of ions traversing the droplet along a radial trajectory. The dependence is shown for ion charges $4e$, $5e$, and $6e$.}
\end{centering}
\end{figure}

%The hope is that such an explosion can be observed.
In order for such an explosion to be detectable, the radius $R$ should be sufficiently large.
%\footnote{The conservative estimate yields the smallest radius for a given $S_e$ and other parameters.};
%the stopping power $S_e$ is a critical parameter for this in Eq.~(\ref{condition}).
Its maximum value $R_{max}$, is obtained from~(\ref{condition}) as,
\begin{eqnarray}
R_{max}=\frac{1}{\gamma+1} \frac{\beta^4}{4\sigma}{S_e}~, \label{rmax}
\end{eqnarray}
and it is linear with respect to the stopping power $S_e$. At a room temperature of 293~K ($\sigma=7.28\times 10^{-2}$~N/m), the dependencies of $R_{max}$ on the ion's energy at different charges of carbon ion are shown in Fig.~\ref{fig:RE}. For the same ions, the maximum radius can be increased by decreasing the surface tension, which can be achieved by increasing the temperature and/or choosing a liquid with smaller surface tension, such as methanol or ethanol.

%Condition~(\ref{condition}) for the observation of a droplet explosion can be rewritten as a condition for the charge of the ion,
%\begin{eqnarray}
%z~>~6 \sqrt{\frac{4(\gamma+1)\sigma}{\beta^4}\frac{R}{S_e(z=6)}}.\label{imp1}
%\end{eqnarray}
%This gives us an idea of how heavier ions can be used for such an observation. The dependence of $z_{min}$ on $R$ at $E=0.5$~MeV/u is shown in

Thus, the key idea is that if the high-LET ion passing through a liquid droplet does induce the shock wave, this wave makes the droplet disintegrate if the pressure at the wave front is larger than the surface tension at the droplet's surface. Now, we need to consider other effects that limit this effect from different sides.

\subsection{Collapse of the cavity in the wake of the shock wave}

However, the opposite effect, the collapse of the cylindrical cavity that formed in the wake of shock wave is expected to follow. First, when this cavity is formed, the inner side is still very hot, and the surface tension pressure on the inner surface is zero. Then, as the expansion continues (for several picoseconds), the surface cools down and the surface tension pressure increases. This pressure on the cylindrical surface is $\sigma/R_{in}$ and since $R_{in}$ is smaller than the radius of the wave front, there is a chance that this pressure is larger than that of the pressure on the wave front. If this is the case, the outgoing pressure wave can slow down and the flow may be inverted to fill the cavity. Pressure oscillations follow this process, but their strength can only be weaker than that of the original shock wave.

Nevertheless, if the shock wave, strong enough to break the droplet's surface tension, reaches the outer surface before the surface tension on the inner surface is developed, the droplet is expected to disintegrate, since the inner layers of the droplet will be slowing down while the outer will still be accelerating in the outward direction.

Let a cylindrical shock wave propagate in the droplet for some time $t$, just sufficient for the inner surface to cool down. Then the condition for stopping the advance of the wave front is:
\begin{eqnarray}
\frac{1}{\gamma+1} \frac{\beta^4}{2}\frac{S_e}{r^2}2\pi r l < \frac{\sigma}{R_{in}}2\pi R_{in} l~.
\label{innercondition1}
\end{eqnarray}
This condition simply represents the Newton's second law, manifesting that the net force acting on the cylindrical layer of the droplet is directed inside; the corresponding multipliers $2\pi r l$ and $2\pi R_{in} l$ on both sides are the outer and inner surface areas that replace the corresponding pressures for forces. Evidently, $R_{in}$ on the r.h.s. of this inequality cancel out and the condition becomes,
\begin{eqnarray}
\frac{1}{\gamma+1} \frac{\beta^4}{2}\frac{S_e}{\sigma} < r ~.
\label{innercondition2}
\end{eqnarray}
Comparing this inequality to~(\ref{rmax}), we can write that it is equivalent to $r>2 R_{max}$, i.e., the cylindrical layer expanding due to a pressure wave will start slowing down as a result of action of the surface tension on the inner surface, {\em independently} of the radius and, therefore, of the time the inner surface cools down, when the radius of the wave front is twice the maximum radius of the droplet. This means that if we choose the maximum radius of the droplet according to Eq.~(\ref{rmax}), the mending effect will {\em not} stop the disintegration of the droplet.

\subsection{Rayleigh instability}

In previous sections, we discussed the upper limit for the radius of a droplet that is deemed to explode because of a shock wave. There is also a lower limit to the droplet's radius. This first limitation is due to the Rayleigh instability~\cite{Rayleigh} brought about by the charge of a droplet. Before the ion enters a droplet it is electrically neutral, but after a traverse, it acquires a charge. If this charge is large enough for a given radius, the droplet may disintegrate because of the Coulomb repulsion rather than the shock wave.

The charge of the droplet, $Ze$, is a sum of two sources. First, when the ion propagates in the medium, it may pick-off electrons. The cross section for this process, known as charge transfer strongly depends on the velocity as ions slow down. Estimates calculated using {Refs.~\cite{Barkas63,Liamsuwan} suggest that for carbon ions the charge transfer becomes significant when the kinetic energy of the ion is smaller than 0.2~MeV/u.} Therefore, in the following examples, 0.5~MeV/u ions are used in order to minimise this effect (since a diminishing value of $z$ decreases the LET) and simplify the estimates.
Second, some electrons ejected from molecules of the medium may be able to escape the droplet and leave it positively charged. Important, that this happens within 50~fs after the ion's traverse, i.e., before the formation of the shock wave~\cite{Radicals}. While the first effect is limited by the number of vacancies in the ion ($z$), the second effect may be considerably stronger, since, e.g., for 0.5-MeV/u C$^{6+}$ ions, about 20 secondary electrons are ejected from every nm of the path~\cite{MSAColl}.

The second effect is limited by the range of propagation of secondary electrons in a medium. If the radius of a droplet is sufficiently large, only a small fraction of ejected electrons escape from it. Equation~(\ref{rmax}) estimates radii of droplets to be traversed by carbon ions to be of the order of several hundreds of nm, while most of the ejected electrons will have energies below 50~eV~\cite{MSAColl} and they are not likely to propagate further than 3~nm~\cite{Radicals}. Only a few $\delta$-electrons will have ranges up to 50~nm~\cite{Meesungnoen02}, but their effect on the charge of droplets is going to be negligible, since the production of $\delta$-electrons at the considered energies (close to the Bragg peak) is suppressed~\cite{MSAColl}.

The estimate for the maximum radius, at which the Coulomb forces overcome the surface tension can be found similarly to Ref.~\cite{Lyalin}, where the surface tension energy is simply compared to the Coulomb energy,
\begin{eqnarray}
\sigma 4\pi R^2 > \frac{Z^2e^2}{2R}~. \label{coulomb}
\end{eqnarray}
The estimate of charge $Z$ in this formula can be obtained if we assume that all electrons ejected from the ion's path within the range-distance from ion's exit or entrance escape the droplet, i.e., $Z=2\frac{dN}{dx}s$, where $s=5$~nm is the range of electrons taken for this (conservative) estimate.
%The estimate includes an assumption that the charges in a droplet are distributed uniformly. This assumption is justified by the timing, i.e., the time scale is so short, that the electrons are just ``blown away'' from the vicinity of the ion's path and the rest of the dynamics is not yet developed. However, this coefficient does not really matter and a more elaborated estimate is not going to be much different. If we maximise $Z$ by assuming that all secondary electrons escape, $Z=\frac{dN}{dx}2R$, we obtain,
%\begin{eqnarray}
%R > \left(\frac{dN}{dx}\right)^2\frac{3e^2}{5\pi\sigma}~. \label{coulomb2}
%\end{eqnarray}
%This limit (shown in Figs.~\ref{fig:RRZ}-\ref{fig:RRZx}) requires further analysis. For 0.5-MeV carbon ion with $z=6$, this estimate gives 250~nm, i.e., that if all secondary electrons escape the droplet, it has to be larger than this size to survive the Coulomb explosion. However, there is no chance that all secondary electrons escape such a droplet, only a small fractions near the ion's entrance and exit can do so. Thus, the actual estimate for the minimal radius is much smaller and is related to the range of secondary electrons in liquid. A better estimate can be obtained if we assume that all electrons ejected from the ion's path within the range-distance from ion's exit or entrance escape the droplet. Then (\ref{coulomb2}) is modified to
The estimate for the minimal radius of the droplet given by the Rayleigh instability is then given by,
\begin{eqnarray}
R > \left[\left(\frac{dN}{dx}\right)^2\frac{e^2 s^2}{\pi\sigma}\right]^{1/3}~. \label{coulomb3}
\end{eqnarray}
%where $s$ is the range of electrons. Even if we use an improbable $s=50$~nm, (\ref{coulomb3}) reduces to 86~nm for $z=6$. The more realistic $s=15$~nm, reduces $R_{min}$ to 40~nm.

Besides this estimate, it is worthwhile to discuss the charge distribution inside the droplet by 50~fs, the time when all remaining electrons are nearly thermalised. This can be done using the calculations of secondary electron transport~\cite{Radicals}. By 50~fs after ion's traverse, there is a thin (of less than 1~nm radius) cylinder positively charged due to ions of water and H$_3$O$^+$ that is almost superseded with a distribution of low-energy electrons that are spread over a slightly wider (1-1.5~nm radius cylinder). So, by and large, electrostatic configuration is that of a cylindrical capacitor with an excesses of positive charge close to the cylinder's bases. The ``capacitor'' part undoubtedly carries energy, but an explosion is unlikely due to attractive forces. However, the repulsive forces can create two jets of water molecules coming out from bases of the cylinder. This is even more likely because the temperature within a cylinder at this time is high and these regions can easily evaporate. However, these jets do not destroy the droplet. If the surface tension is large enough, the cylinder will be filled in by a nanosecond time.

\begin{figure}
\begin{centering}
\resizebox{0.99\columnwidth}{!}
{\includegraphics{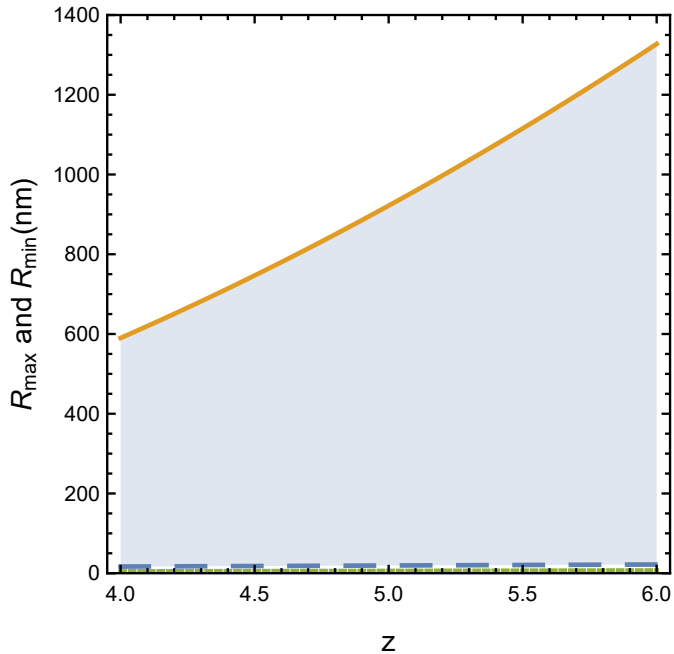}}
\caption{\label{fig:RRZ}The dependence of the maximal (solid line), calculated using (\ref{rmax}), and minimal radii (dashed line), calculated using (\ref{coulomb3}), of the liquid water droplet (at room temperature), at which its explosion due to the effect of the shock wave is predicted, on the charge $z$ (in units of $e$) of carbon ion traversing the droplet at energy 0.5~MeV/u. The droplets with radii within the shaded area are predicted to explode due to the action of shock waves. The dotted line indicates the minimum radius from the point of avoiding the evaporation of the droplet.}
\end{centering}
\end{figure}

The charge transfer effect makes a tiny correction to this effect; for instance, in the case of carbon ions, an increase of charge by the maximum value of $6e$ {is tiny compared to the charge of the order of $200$, obtained from the above estimate.} Another effect of forward and backward emitted electron jets, reported in Ref.~\cite{Boecking} may also affect the charge remaining on a droplet, but it is unlikely to change the picture significantly. The results of the estimates of $R_{min}$ using~(\ref{coulomb3}) and $R_{max}$ using~(\ref{rmax}) for 0.5~MeV/u ions with different charges are shown in Fig.~\ref{fig:RRZ}.
\begin{figure}
\begin{centering}
\resizebox{0.99\columnwidth}{!}
{\includegraphics{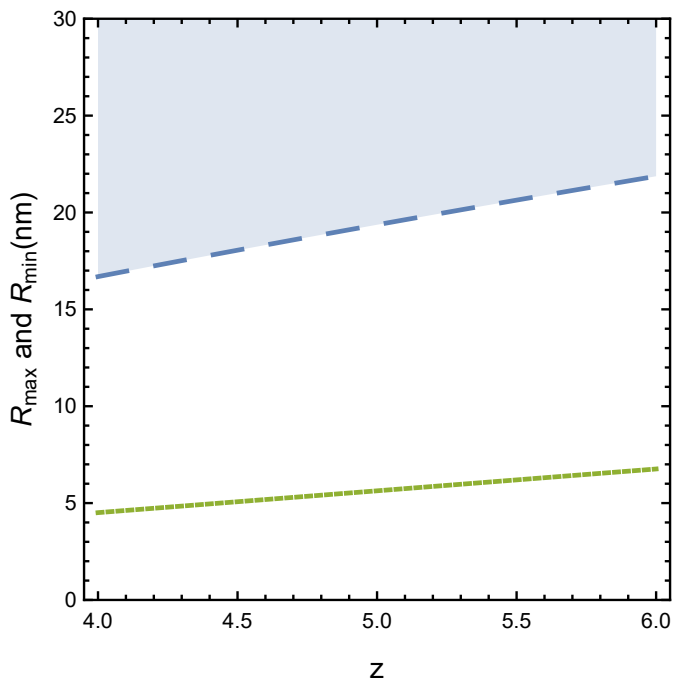}}
\caption{\label{fig:RRZx} A detail of Fig. \ref{fig:RRZ}, showing the minimal radii in more detail.}
\end{centering}
\end{figure}
These results tell us that the effect of Rayleigh instability cannot cause the multifragmentation of a droplet with a radius larger than 25~nm.

\subsection{Evaporation of the droplet}
\label{sec.ev}

Certainly, if too much energy is deposited to a droplet, it can evaporate. Then the shock wave cannot be a culprit of its ``disintegration''. A conservative estimate for this effect can be done rather easily. The maximal energy deposited in the droplet is equal to $S_e 2R$; this energy must not exceed the heat needed to evaporate the droplet, $\frac{4}{3}\pi R^3 \varrho (\lambda+c_p \Delta T)$, where $\lambda$ is the latent heat of vaporisation of liquid, $c_p$ is its specific heat, and $\Delta T$ is the temperature difference between the temperature of vaporization and the temperature of the droplet before the traverse. This gives the following limitation to the radius of the droplet:
\begin{eqnarray}
R>\sqrt{\frac{3 S_e}{2\pi \varrho (\lambda+c_p \Delta T)}}~.
\label{evaporation}
\end{eqnarray}
The results for this limitation are shown in Figs.~\ref{fig:RRZ}-\ref{fig:RRZx} with a dotted line. It appears that this limitation is not at all stringent for the experiments that we can think of, but it is surely important for the molecular dynamics simulations of this process. { In parallel with this work, the classical molecular dynamics (MD) simulations have been conducted using the MBN Explorer package~\cite{MBNX}. The largest droplet thus far modelled is of radius 10~nm. It contains about 420 thousand atoms and is already quite computationally expensive. For such a droplet, the window between evaporating droplet and a too weak shock wave is very narrow. The results of these simulations will be presented in a separate paper.}

{However, the evaporation of water droplets in vacuum (where they are supposed to be irradiated) may be significant for the observations. It is not related to irradiation of droplets, but to the fact that lifetime of droplets in vacuum is limited, i.e., they may evaporate before they have a chance to be irradiated. The characteristic time of multifragmentation due to action of shock wave is 10-100~ps, depending on the size of a droplet. The characteristic time of evaporation of a droplet can be estimated as
\begin{eqnarray}
\frac{R}{v}\exp{\frac{\lambda_0}{kT}}~,
\label{evaporation2}
\end{eqnarray}
where $v$ is an average velocity of a water molecule, $\lambda_0$ is a latent heat of vaporisation per molecule, and $k$ is Boltzmann constant. For a 100-nm radius droplet, this estimate gives $10^{-4}$~s. This time is five orders of magnitude longer than the time of the discussed multifragmentation, which makes us hopeful that the the observation of the latter is possible.
}

%%%%%%%%%%%%%%%%%%%%%%%%%%%%%%%%%%%%%%%%%%%%%%%%%%%%%%%%%%%%%%%%%

\section*{Conclusion}

Thus, if an ion enters a droplet with a radius limited by inequality~(\ref{condition}), but larger than obtained from estimate~(\ref{coulomb3}), we predict that the droplet will disintegrate into smaller parts due to the action of the shock wave. If the collision is not central, a disintegration happens more readily. As an example, let us consider a droplet traversed by a fully charged ($z=6$) carbon ion at ${\cal E}=0.5$~MeV/u. Its stopping power given by~(\ref{bethe}), is $\approx1.5$~keV/nm. According to Fig.~\ref{fig:RRZ}, an explosion of the droplet due to a shock wave will happen if its radius is somewhere between 30 and 1000~nm.
%{\color{red} Let us take $R=400$~nm.}
If the ion passes some 800~nm inside the droplet, it loses about 1.2~MeV or 0.1~MeV/u. Therefore, we can safely neglect the change in the ion's speed during the traverse. At this energy, the mean free path related to charge transfer process is larger than 200~nm~\cite{NikjooCTran}. Therefore, even if this ion picks off an electron or even two, all criteria are satisfied (Fig.~\ref{fig:RRZ}) and we predict that the droplet will disintegrate.

In absence of shock waves, there is no mechanism for a droplet to explode and it will evaporate within $10^{-4}-10^{-3}$s, as explained in section~\ref{sec.ev}. Indeed, if the diffusion and heat conductivity are the only mechanisms for heat transfer, as discussed in Refs.~\cite{preheat,Radicals}, then the pressure at a distance of 20~nm (for carbon ions) will be close to the atmospheric at 1~ps, and will hardly increase at later times. The temperature of, e.g., a 400-nm-radius droplet will increase by less than 1~K and the droplet will surely survive.
%Thus, the observation of the droplet's explosion proves the existence of shock waves due to ions at high LET propagating in the medium, while the failure to observe such explosions refutes their existence.

\begin{figure}
\begin{centering}
\resizebox{0.95\columnwidth}{!}
{\includegraphics{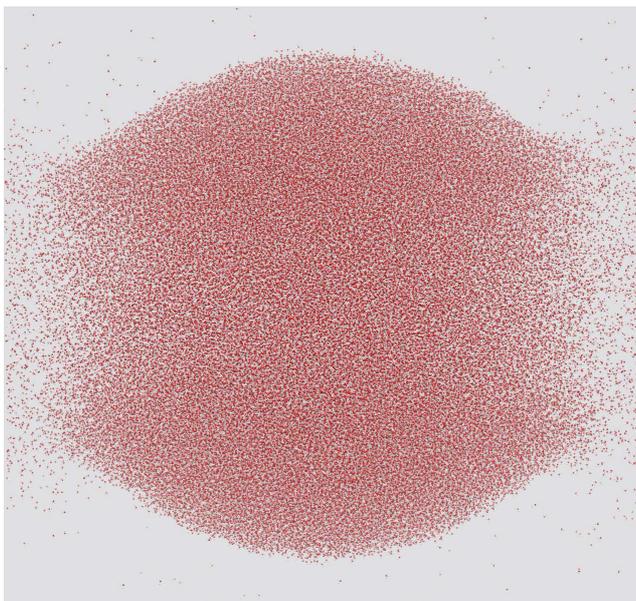}}
\caption{\label{fig:art} A snapshot of a disintegrating droplet obtained from MD simulations using MBN Explorer package. The ion's path is horizontal. The snapshot is done at about 2~ps time after the onset of the shock wave.}
\end{centering}
\end{figure}

We suggest an experiment to observe predicted shock waves on a nanometre scale initiated by ions with energy $0.1-0.5$~MeV/u propagating in tissue-like media, such as liquid water. Figure~\ref{fig:art} gives a snapshot of MD simulation of multifragmentation of a liquid droplet traversed by an ion.
These simulations are only the first step in this study and we hope that such simulations will help us to learn more about the dynamics of multifragmentation of a droplet. We hope that such an experiment is carried out, since the discovery of shock waves induced by ions is critical for scientific fundamentals of ion-beam therapy and understanding the radiation damage induced by ions.
%An ion's collision inside a droplet constitutes a critical event. The outcome of such events with respect to survival or explosion of a droplet will indicate the existence or non-existence of such shock waves.

\section*{Acknowledgements}

It was a pleasure to discuss this paper with Gleb Gribakin in Belfast, UK. We appreciate the support of FP7 ITN-ARGENT. ES is indebted to G.~Sushko and P. de Vera for their assistance in modeling shock waves using the MBN Explorer package.

\bibliographystyle{epj}
%\bibliography{bibliography1}

\end{document}